\documentstyle[prl,aps]{revtex}

\begin{document}
\draft

\twocolumn[\hsize\textwidth\columnwidth\hsize\csname@twocolumnfalse\endcsname

\title{
Stability of quantum states of
finite macroscopic systems against\\
classical noises, perturbations from environments, 
and local measurements
}

\author{Akira Shimizu\cite{shmz} and Takayuki Miyadera\cite{MIYA}
}

\address{
Department of Basic Science, University of Tokyo, 
3-8-1 Komaba, Tokyo 153-8902, Japan}


\maketitle

\begin{abstract}
We study the stability of quantum states of macroscopic systems of
finite volume $V$, 
against weak classical noises (WCNs), 
weak perturbations from environments (WPEs), 
and local measurements (LMs).
We say that a pure state is `fragile' if its decoherence rate is 
anomalously great, and `stable against LMs' if
the result of a LM is not affected by another LM at a distant point.
By making full use of the
locality and huge degrees of freedom, we show the following:
(i) If square fluctuation of {\em every} additive operator is $O(V)$
or less for a pure state, then it is not fragile in {\em any} 
WCNs or WPEs.
(ii) If square fluctuations of {\em some} additive operators are $O(V^2)$
for a pure state, then it is fragile in {\em some} 
WCNs or WPEs.
(iii) If a state (pure or mixed) has the `cluster property,'
then it is stable against LMs, and vice versa.
These results have many applications, among which 
we discuss the mechanism of symmetry breaking in finite systems.
\end{abstract}
\pacs{PACS numbers: 03.65.Yz, 03.65.Ta, 11.30.Qc, 05.40.Ca}
]

The stability of quantum states of macroscopic systems, 
which are subject to weak classical noises (WCNs) 
or weak perturbations from environments (WPEs), have been 
studied in many fields of physics as the decoherence problem \cite{dec}.
However, most previous studies assumed
that the principal system was describable 
by a small number of collective coordinates.
Although such models might be applicable to 
some systems,
applicability to general systems is questionable.
As a result of the use of such models, 
the results depended strongly on the choices of 
the coordinates and the form of the interaction $\hat H_{\rm int}$
between the principal system and a noise or an environment \cite{dec}.
For example, a robust state for some $\hat H_{\rm int}$ can become
a fragile state for another $\hat H_{\rm int}$.
However, macroscopic physics and experiences strongly indicate that 
a more universal result should be drawn.

In this paper, we study 
the stability of quantum states of finite 
macroscopic systems against WCNs and WPEs.
We also propose a new criterion of stability; the stability against  
local measurements (LMs).
We study these stabilities using 
a general model with a macroscopic number of degrees of 
freedom $N$.
In addition to the fact that $N$ is huge, we make 
full use of the locality 
\cite{haag,ft} ---
`additive' observables must be 
the sum of local observables over a macroscopic region,
the interaction $\hat H_{\rm int}$ must be local,
and measurement must be local.
By noticing these points, 
we derive general and universal results.
Among many applications of the present theory, 
we discuss the mechanism of symmetry breaking 
in finite systems.

{\em Macroscopic quantum systems:}
As usual, we are only interested in phenomena in some energy 
range $\Delta E$,
and describe the system 
by an effective theory which correctly describes the system only 
in $\Delta E$.
For a given $\Delta E$, 
let ${\cal M}$ be the number of {\em many-body} quantum states in that 
energy range.
Then, 
\[
N \sim \ln {\cal M}
\]
is the degrees of freedom of the effective theory.
Note that $N$
can become a small number 
even for a system of many degrees of freedom
when, e.g., a non-negligible energy gap exists in $\Delta E$, 
as in the cases of 
a heavy atom at a meV or lower energy range
and SQUID systems at low temperatures.
{\em We here exclude such systems}, 
because they are essentially systems of small degrees of freedom.
Namely, we say that a system is macroscopic (for a given $\Delta E$)
only when its $N$ is a macroscopic number.
We further assume that the system extends homogeneously \cite{homogeneous}
over a volume $V$, and that boundary 
effects are negligible.
Since $\Delta E$ sets 
a minimum length scale $\ell$, 
\[
V \sim N \ell^d
\]
in $d$ dimension.
We therefore say that $V$ is also macroscopic.
We study the stability of states of such a macroscopic 
system, when it is subject to WCNs, WPEs, and LMs.
The Hamiltonian $\hat H$ of the system
can be a general one which has only short-range interactions.

{\em Measures of the correlations between distant points:}
As we shall show later, 
correlations between distant points are important. 
As a measure of the correlations, we first consider the 
cluster property (CP). 
In {\em infinite} systems, 
a quantum state is said to have the CP
if 
\[
\langle \delta \hat a(x) \delta \hat b(y) \rangle
\to 0
\mbox{ as } |x - y| \to \infty
\]
for every local operators $\hat a(x)$ and $\hat b(y)$
at $x$ and $y$, respectively, 
where 
$
\delta \hat a(x) \equiv \hat a(x) - \langle \hat a(x) \rangle
$
and 
$
\delta \hat b(y) \equiv \hat b(y) - \langle \hat b(y) \rangle
$ \cite{confuse}.
%
Here, by a {\em local operator at $x$} 
we mean a finite-order polynomial of 
field operators and their finite-order derivatives at position $x$
\cite{ft}.
We generalize the concept of the CP to
the case of finite systems as follows\cite{MS}.
For a small positive number $\epsilon$, 
we define a region $\Omega(\epsilon,x)$
by its complement 
$\Omega(\epsilon,x)^c$, which is the region of $y$ in which 
\begin{equation}
\left|
\langle \delta \hat a(x) \delta \hat b(y) \rangle
\right|
\leq
\epsilon
\sqrt{
\langle \delta \hat a^\dagger(x) \delta \hat a(x) \rangle
\langle \delta \hat b^\dagger(y) \delta \hat b(y) \rangle
}
\label{CP}\end{equation}
for every local operators $\hat a(x)$ and $\hat b(y)$.
Let 
$\Omega(\epsilon) \equiv \sup_x |\Omega(\epsilon,x)|$, 
where $|\Omega(\epsilon,x)|$ denotes the size of $\Omega(\epsilon,x)$.
Intuitively, $\Omega(\epsilon)$ is the size of the region
outside which correlations of every local operators becomes negligible.
We consider a sequence of homogeneous \cite{homogeneous} systems
with various values of $V$ and associated states, 
where the shapes of $V$'s are similar to each other.
(For example, the ground states of many particles in 
spherical boxes with various sizes, 
with the same particle density.)
We say that the states (for large $V$) of the sequence
have the CP 
if $\Omega(\epsilon)$ for any $\epsilon > 0$ becomes
independent of $V$ for a sufficiently large $V$.
This means that 
$\Omega(\epsilon) \ll V$
if one takes $V$ large enough.
Note that a small number of Bell pairs do not destroy 
the CP: the lack of the CP means a {\em macroscopic} entanglement.

As a second measure, 
we consider fluctuations of additive quantities. 
A physical quantity $A$ is `additive' if
\[
A = A^{(1)} + A^{(2)}
\]
when we regard the system as a composite system of subsystems 1 and 2.
Thermodynamics {\em assumes} that 
any states in a pure phase satisfies
$
\langle (\delta A)^2 \rangle = o(V^2)
$
for {\em every} additive quantity. 
In particular, 
if a state of (quantum or classical) system satisfies
\[
\langle (\delta A)^2 \rangle \leq O(V)
\]
for {\em every} additive quantity, 
we call it a `normally-fluctuating state' (NFS).
In {\em finite} quantum systems, 
on the other hand, 
there exist pure states for which {\em some} of
additive operators have anomalously-large fluctuations;
\[
\langle (\delta \hat A)^2 \rangle
= O(V^2).
\]
We call such a {\em pure} state an `anomalously-fluctuating state' (AFS).
The locality requires that 
additive operators of quantum systems must have the following form:
\[
\hat A
=
\sum_{x \in V} \hat a(x),
\]
where $\hat a(x)$ denotes a local operator at $x$.
It is easy to show that 
an AFS does not have the CP,  hence is entangled macroscopically.
For {\em infinite} quantum systems,  
there is a well-known theorem: 
{\em Any pure state has the CP} \cite{haag}.
Therefore, AFSs 
converge (in the weak topology) into {\em mixed} states 
as $V \to \infty$, although they are pure states in finite systems \cite{ex}.
Since AFSs are such unusual states, they are expected to be
unstable in some sense.
We now clarify in what sense, how, and why unstable.

{\em Fragility:}
We say a quantum state is `fragile' if
its decoherence rate $\Gamma$ (see Eq.\ (\ref{Gamma}))
behaves as 
\[
\Gamma \sim 
K V^{1+\delta}
,\]
where $K$ is a function of microscopic parameters, 
and $\delta$ is a positive constant.
To understand the meaning of the fragility, 
consider first the non-fragile case where $\delta=0$.
In this case, $\Gamma/V$ 
is independent of $V$. 
This is a normal situation in the sense that 
the total decoherence rate $\Gamma$ is 
basically the sum of {\em local} decoherence rates,
which are determined only by 
microscopic parameters. 
%
On the other hand, 
the case $\delta > 0$ is an anomalous situation in which 
$
\Gamma/V \sim K V^{\delta}$.
Note that this can be very large even when $K$ is small,
because, by definition, a macroscopic volume is huge.
This means that
{\it a fragile quantum state 
decoheres due to a noise or environment 
at an anomalously great rate}, 
even when the coupling constant between 
the system and the noise or environment is small.

{\em Fragility under WCN:}
The point of the present theory 
is the locality \cite{haag}.
For the Hamiltonian $\hat H_{\rm int}$ of the interaction 
with a classical noise, 
the locality requires that it 
should be the sum of local interactions \cite{general};
\begin{equation}
\hat H_{\rm int}
=
\lambda
\sum_{x \in V}
f(x,t) \hat a(x).
\label{int-noise}\end{equation}
Here, 
$\lambda$ is a small positive constant, 
$f(x,t)$ is a random classical noise field with vanishing average
$\overline{f(x,t)} = 0$, 
and $\hat a(x)$ is a local operator at $x$.
We assume that 
$\overline{f(x,t) f(x',t')}$ depends only on
$x-x'$ and $t-t'$, 
and that its correlation time 
$\tau_{\rm c} \ll 1/\Gamma$ \cite{time}.
We denote the spectral intensity of $f$ by $g(k,\omega)$ \cite{k}, which is 
positive by definition.
%
A pure state $| \Psi \rangle$ at $t=0$ evolves for $t>0$ 
by the total Hamiltonian
$ 
\hat H + \hat H_{\rm int}
$,
and 
the density operator is given by

\[
\hat \rho(t) \equiv \overline{| \Psi(t) \rangle \langle \Psi(t) |}.
\]
Since we are interested in the dependence of $\Gamma$ on the initial state,
we study 
an early time stage $\tau_{\rm c} \ll t \ll 1/\Gamma$, 
and define $\Gamma$ as the increase rate 
of the $\alpha$-entropy 
of $\alpha=2$ in this time region;
\begin{equation}
\Gamma 
\equiv
- \frac{1}{2} \frac{d}{d t} \ln {\rm Tr} [\hat \rho(t)^2]
\Big|_{\tau_{\rm c} \ll t \ll 1/\Gamma}.
\label{Gamma}\end{equation}
Even when $\hat H_{\rm int} = 0$, $| \Psi \rangle$ generally evolves 
by $\hat H$. 
Since we are interested in the instability induced by $\hat H_{\rm int}$, 
we consider states which do not evolve by $\hat H$
in this time region, 
i.e.,  
$\exp(-i \hat H t)| \Psi \rangle 
\simeq \exp(-i \langle \hat H \rangle t)| \Psi \rangle$ for such $t$.
(However, see \cite{relax}.)
Moreover, since we are interested in
the case of weak noise, 
we evaluate $\Gamma$ to $O(\lambda^2)$.
By dropping non-dissipative contributions from 
$\hat H_{\rm int}$,
because they can be absorbed in $\hat H$ as renormalization terms,
we find
\begin{eqnarray*}
\Gamma
&\simeq&
\lambda^2 \sum_{k,n}
g(k, \langle \hat H \rangle - \omega_{n})
|
\langle n | \delta \hat A_k | \Psi \rangle
|^2
\\
&\equiv&
\lambda^2 \sum_{k}
g(k)
\sum_{n}
|
\langle n | \delta \hat A_k | \Psi \rangle
|^2,
\end{eqnarray*}
where 
we have defined $g(k)$ by the last equality.
Here, 
$| n \rangle$ is an eigenstate of $\hat H$,
with eigenenergy $\omega_n$ (which may be degenerate), 
and
$
\delta \hat A_k
\equiv
\hat A_k - \langle \Psi | \hat A_k | \Psi \rangle
$,
where
\[
\hat A_k
\equiv
\sum_{x \in V} \hat a(x) e^{-i k x}
.\]
Since 
$g(k, \langle \hat H \rangle - \omega_{n})$ and 
$|\langle n | \delta \hat A_k | \Psi \rangle|^2$ 
are both positive, 
$g(k)$ may be interpreted as
a typical (average) value of $g(k, \langle \hat H \rangle - \omega_{n})$
for relevant $n$'s.
This interpretation would be good at least 
for the $V$ dependence, which is of our primary interest. 
We then obtain the simple formula;
\begin{equation}
\Gamma
\simeq
\lambda^2 \sum_k
g(k)
\langle \Psi | \delta \hat A_k^\dagger \delta \hat A_k | \Psi \rangle.
\label{r-noise-2}\end{equation}
Note that $\hat A_k$ is an additive operator
because $\hat a(x) e^{-i k x}$ is a local operator.
When $\hat a(x)$ is a spin operator, 
e.g., 
$\hat A_k$ for $k = \pi/\ell$ is 
the staggered magnetization.

When $| \Psi \rangle$ is an NFS, 
$\langle \Psi | \delta \hat A_k^\dagger \delta \hat A_k | \Psi \rangle
\leq O(V)$
for any $\hat A_k$, hence
\[
\Gamma
\lesssim \lambda^2
O(V) \sum_k g(k)
.\]
Since 
\[
\sum_k g(k,\omega_n - \omega_{n'})
= \int \overline{f(x,t)f(x,0)} e^{i (\omega_n - \omega_{n'}) t} dt
\]
does not depend on $V$, 
neither $\sum_k g(k)$ does.
We thus find that 
{\em NFSs are not fragile in any WCN}.
When $| \Psi \rangle$ is an AFS, 
on the other hand, 
$\langle \Psi | \delta \hat A_k^\dagger \delta \hat A_k | \Psi \rangle
= O(V^2)$
for some $\hat A_k$, i.e., for some $\hat a(x)$ and some $k=k_0$.
Hence, if $\hat H_{\rm int}$ has a term that 
is composed of such $\hat a(x)$'s, then
\begin{equation}
\Gamma
\simeq
\lambda^2 O(V^2) g(k_0)+
\lambda^2 O(V) \sum_{k \neq k_0} g(k),
\label{r-noise-AFS}\end{equation}
and the AFS becomes fragile if 
$g(k_0) = O(V^{-1+\delta})$, where $\delta >0$.
Therefore, {\em an AFS is fragile in some WCN.}

{\em Fragility under WPE:}
The physical realities of noises are 
perturbations from environments.
We obtain similar results for WPEs. 
From the locality, the interaction with an environment
should be the sum of local interactions \cite{general};
\[
\hat H_{\rm int}
=
\lambda \sum_{x \in V}
\hat f(x) \hat a(x)
.\]
Here, $\hat f(x)$ and $\hat a(x)$ are local operators at $x$ of
an environment and the principal system, respectively. 
For $\langle \hat f(x,t) \rangle_{\rm E}$ and 
$\langle \hat f(x,t) \hat f(x',t') \rangle_{\rm E}$
(in the interaction picture), 
where $\langle \cdots \rangle_{\rm E}$ denotes
the expectation value for the state $\hat \rho_{\rm E}$
of the environment E, 
we assume the same properties as
$\overline{f(x,t)}$ and 
$\overline{f(x,t) f(x',t')}$, respectively, of the WCN.
The total Hamiltonian is 
\[
\hat H + \hat H_{\rm int} + \hat H_{\rm E},
\]
where $\hat H_{\rm E}$ is the Hamiltonian of E.
Taking the initial state $\hat \rho_{\rm total}(0)$ as the product state
$
| \Psi \rangle \langle \Psi | \otimes \hat \rho_{\rm E}
$, 
we evaluate the reduced density operator
\[
\hat \rho(t) \equiv {\rm Tr_E} [\hat \rho_{\rm total}(t)]
.\]
%
We then obtain the same result (\ref{r-noise-2}), 
where $g(k,\omega)$ is now the spectral intensity derived from 
$\langle \hat f(x,t) \hat f(x',t') \rangle_{\rm E}$.
Therefore, 
{\em NFSs are not fragile under any WPE},
while 
{\em 
AFS are fragile under some WPE
}\cite{ekert}.

{\em Summary of fragility:}
We have shown
that NFSs are not fragile in any WCNs or WPEs.
This should be contrasted with the results of most previous works, 
according to which a state could be either fragile or robust 
depending on the form of $\hat H_{\rm int}$ \cite{dec}.
Note that our results concern an 
{\em approximate} stability (i.e., non-fragility)
against {\em all} possible WCNs or WPEs and $\hat H_{\rm int}$'s, 
whereas most previous works studied 
the {\em exact} stability against {\em particular} ones. 
We think that the former is more important in macroscopic systems
because
many types of WCNs or WPEs and $\hat H_{\rm int}$'s 
would coexist in real systems, 
and the exact stability against some of them could not exclude
fragility to another.
%
Regarding AFSs, on the other hand, our results show only that
they are fragile in {\em some} WCN or WPE.
In other words, 
for any AFS
it is always possible to {\em construct} 
a noise (or an environment) and a weak local interaction with it 
in such a way that the AFS becomes fragile.
These results do {\em not} guarantee the existence
of the relevant noise (or an environment) and the relevant interaction
{\em in real physical systems}.
Since there is no theory that is general enough on
WCNs or WPEs at present, we cannot draw a definite 
conclusion on whether AFSs are {\em always} fragile in 
real physical systems. 
It rather seems that, as we will discuss later, there may be some cases 
where some AFSs are non-fragile, 
in contradiction to naive expectations.
This motivates us to explore the following new stability.

{\em Stability against LMs:}
Suppose that one performs an ideal (von Neumann) measurement of a local 
observable $\hat a(x)$  at $t=t_a$
for a state $\hat \rho$ (pure or mixed) of a macroscopic system, 
and obtains a value $a$ with a finite probability $P(a) \neq 0$.
Subsequently, one measures another local 
observable $\hat b(y)$ at a later time $t_b$ \cite{like},
and obtains a value $b$.
Let $P(b;a)$ be 
the probability that $b$ is obtained at $t_b$ under the condition
that $a$ was obtained at $t_a$.
On the other hand, 
one can measure $\hat b(y)$ at $t=t_b$ without performing 
the measurement of $\hat a(x)$ at $t_a$.
Let $P(b)$ be the probability distribution of $b$ in this case.
We say $\hat \rho$ is `stable against local measurements' if
for any $\varepsilon >0$
\begin{equation}
\left| P(b;a) - P(b) \right| \leq \varepsilon
\
\mbox{for sufficiently large $|x-y|$,}
\label{SLM}\end{equation}
for {\em any} local operators $\hat a(x)$ and $\hat b(y)$
and their eigenvalues $a$ and $b$ such that 
$P(a) \geq \varepsilon$.
For the simplest case $t_b \to t_a$, 
we obtain 
the simple theorem:
{\em If $\hat \rho$ is stable against LMs then 
it has the CP, and that 
any state which has the CP is stable against LMs.}
It follows, e.g., that
any AFS is unstable against LMs.

To prove this theorem, 
we use the spectral decomposition \cite{continuous};
\[
\hat a(x) = \sum_a a \hat{\cal P}_a(x)
,\]
and similarly for $\hat b(y)$.
Here, $\hat{\cal P}_a(x)$ 
denotes the projection operator 
corresponding to an eigenvalue $a$ of $\hat a(x)$.
Note that 
$[\hat{\cal P}_a(x), \hat{\cal P}_b(y)] = 0$
for $t_b \to t_a$.
Since we are considering an effective theory in a finite energy range, 
we assume that ultraviolet divergences are absent:
e.g., for any positive integer $m$,
\[
\langle \hat a(x)^m \rangle 
= 
\mbox{finite}
\]
for any local operator $\hat a(x)$.
For $t_b \to t_a$, 
both  $| P(b;a) - P(b) | \leq \varepsilon$ and
$| P(a;b) - P(a) | \leq \varepsilon$ are satisfied
if $\hat \rho$ is stable against LMs.
Expressing the probabilities by the projection operators, 
we obtain 
\begin{eqnarray*}
&&
\left|
{\rm Tr} [\hat \rho \hat{\cal P}_a(x) \hat{\cal P}_b(y)]
-
{\rm Tr}[\hat \rho \hat{\cal P}_a(x)]
{\rm Tr} [\hat \rho \hat{\cal P}_b(y)]
\right|
\\
&& \qquad \leq
\varepsilon 
\min \left(P(a), P(b)\right)
\end{eqnarray*}
for $P(a), P(b) \geq \varepsilon$.
Multiplying this equation by $|ab|$, and summing over $a$ and $b$
such that $P(a), P(b) \geq \varepsilon$,
we obtain
\[
\left|
\langle \delta \hat a_\varepsilon(x) \delta \hat b_\varepsilon(y) \rangle
\right|
\leq
\varepsilon K
.\]
Here, $K$ is a finite positive number, 
which does not depend on $|x-y|$, and
$
\hat a_\varepsilon(x)
\equiv
\sum'_a a \hat{\cal P}_a(x),
$
where $\sum'_a$ denotes the summation over $a$ such that  
$P(a) \geq \varepsilon$, and similarly for $\hat b_\varepsilon(y)$.
By letting $\varepsilon \to 0$ (thus increasing $|x-y|$ accordingly), 
we obtain
\[
|
\langle \delta \hat a_\varepsilon(x) \delta \hat b_\varepsilon(y) \rangle
|
\to
|
\langle \delta \hat a(x) \delta \hat b(y) \rangle
|
\to 0
.\]
It is easy to show the CP from this.
To prove the inverse, 
we take
$\hat a(x) = \hat{\cal P}_a(x)$, $\hat b(y) = \hat{\cal P}_b(y)$.
Then, from (\ref{CP}), 
\begin{eqnarray*}
|
\langle \delta \hat{\cal P}_a(x) \delta \hat{\cal P}_b(y) \rangle
|
&\leq&
\epsilon
\sqrt{
P(a)(1-P(a))P(b)(1-P(b))
}
\\
&\leq&
\epsilon
\sqrt{
P(a)P(b)
}
\end{eqnarray*}
for sufficiently large $|x-y|$.
Dividing this by $P(a)$ yields
\[
| P(b;a) - P(b) | 
\leq \epsilon \sqrt{P(b)/P(a)}
\leq \epsilon \sqrt{1/P(a)}
\leq \sqrt{\epsilon}
\]
for $P(a) \geq \epsilon$
(hence, also for $P(a) \geq \sqrt{\epsilon}$).
We thus obtain the stability against LMs.

{\em Applications:}
The above results have many applications,
including quantum computers with many qubits \cite{us}, 
and non-equilibrium statistical physics.
We here discuss the mechanism of symmetry breaking 
in {\em finite} systems,
which 
has been a long-standing question for the following reasons.
Consider a finite system that will exhibit a symmetry breaking if 
$V$ goes to infinity.
Let $|\Psi \rangle_V$ be a state that 
approaches, as $V \to \infty$, a symmetry-breaking vacuum
$|\Psi \rangle_\infty$ of the infinite system,
in the sense that 
\[
\lim_{V \to \infty} \langle \Psi_V| \hat a(x) |\Psi_V \rangle
=
\langle \Psi_{\infty}|\hat a (x) |\Psi_{\infty} \rangle 
\]
for any local operator $\hat a(x)$.
We call $|\Psi \rangle_V$ for large $V$ a pure-phase vacuum.
It has a macroscopic value 
\[\langle \Psi_V| \hat M |\Psi_V \rangle = O(V)
\]
of an additive order parameter $\hat M$.
In a mean-field approximation, 
pure-phase vacua have the lowest energy.
However, 
it is {\em always} possible 
(see the example below)
to construct a pure state(s) 
that does not
break the symmetry, $\langle \hat M \rangle = 0$, and has
an equal or {\em lower} energy than pure-phase vacua \cite{HL,pre01}.
Although such states cannot be pure in infinite systems, 
they can be pure in finite systems \cite{haag,HL,pre01,prl00}.
When $[\hat H, \hat M] \neq 0$, in particular, 
the exact lowest-energy state 
is generally such a symmetric ground state \cite{HL,pre01}.
To lower the energy of a pure-phase vacuum, 
a symmetry-breaking field is necessary.
However, an appropriate symmetry-breaking field would not always exist 
in real physical systems. For example, the 
symmetry-breaking field for antiferromagnets 
is a static staggered magnetic field, which alters its direction at 
the period exactly twice the lattice constant.  It seems quite unlikely 
that such a field would always exist in laboratories. 

Our results suggest the following new mechanisms of 
symmetry breaking in finite systems.
From the well-known theorem mentioned earlier, 
$|\Psi \rangle_\infty$ has the CP.
Since $|\Psi \rangle_V$ approaches $|\Psi \rangle_\infty$, 
it also has the CP for large $V$.
Hence, pure-phase vacua are not AFSs. 
On the other hand, 
$\langle \delta \hat M^2 \rangle = O(V^2)$ 
for the symmetric ground state because
it is composed primarily of a superposition of
pure-phase vacua with different values of $\langle \hat M \rangle$
\cite{HL,pre01}.
Namely, 
the symmetric ground state is an AFS, 
and thus is fragile in some WCN or WPE. 
Therefore, we expect that a pure-phase vacuum would 
be realized much more easily than the symmetric ground state. 
This mechanism may be called ``environment-induced symmetry breaking,''
a special case of which was discussed
for interacting many-bosons
\cite{prl00}.
For general systems, however, there is one delicate point:
$g(k_0)$ of the relevant WCN or WPE might be 
$O(1/V)$ in some of real systems \cite{intensity}.
Then, Eq.\ (\ref{r-noise-AFS}) yields
$\Gamma = O(V)$, and the symmetric ground state becomes non-fragile.
In such a case, we must consider 
the stability against LMs: %
Even when the symmetric ground state is somehow realized at some time, 
it is changed into another state
when one measures (or, `looks' at) a relevant observable that is 
localized within only a tiny part of the system.
Such drastic 
changes continue by repeating measurements of relevant observables, until
the state becomes a pure-phase vacuum and the symmetry is broken.
This mechanism may be called ``measurement-induced symmetry breaking.''
We conjecture (and confirmed in several examples) that
the number of LMs necessary for reducing an AFS to an NFS
would be much less than $N$.

For example, if we regard the spins of the antiferromagnetic
Ising model as quantum spins, the pure-phase vacua are the N\'{e}el states,
\begin{eqnarray*}
| \Psi_+ \rangle &\equiv& 
| \uparrow \downarrow \uparrow \cdots \downarrow \rangle,
\\
| \Psi_- \rangle &\equiv& 
| \downarrow \uparrow \downarrow \cdots \uparrow \rangle,
\end{eqnarray*}
for which the staggered magnetization
\[
\hat M_\pi \equiv \sum_x e^{i \pi x/\ell} \hat \sigma_z(x)
\]
is the order parameter; 
\[
\langle \Psi_\pm | \hat M_\pi | \Psi_\pm \rangle = \pm V
.\]
On the other hand, 
\[
| \Phi \rangle
\equiv
(| \Psi_+ \rangle + | \Psi_- \rangle)/\sqrt{2}
\]
is an symmetric ground state.
(In this simple model, 
$| \Phi \rangle$ is degenerate with $| \Psi_\pm \rangle$.
In more general models, the symmetric ground state is often the 
unique ground state 
\cite{HL,pre01}.)
It is an AFS because 
$
\langle \Phi | \hat \delta M_\pi^2 | \Phi \rangle = V^2
$.
According to our results, $| \Psi_\pm \rangle$ are stable against LMs.
For example, 
after measurement of $\hat \sigma_x$ of the first spin, 
$| \Psi_+ \rangle$ reduces to 
\[
\frac{1}{\sqrt{2}}(| \uparrow \rangle + | \downarrow  \rangle)
\otimes
| \downarrow \uparrow \cdots \downarrow \rangle
\quad 
\mbox{when $\sigma_x = +1$},
\]
or
\[
\frac{1}{\sqrt{2}}(| \uparrow \rangle - | \downarrow  \rangle)
\otimes
| \downarrow \uparrow \cdots \downarrow \rangle
\quad
\mbox{when $\sigma_x = -1$}.
\]
Hence, the result of subsequent measurement of any spin 
operator at a distant point is not affected at all
by the first measurement.
In contrast, 
$| \Phi \rangle$ are unstable against LMs, i.e., 
against measurement of {\em some} local spin operator. 
Namely, 
if the initial state is $| \Phi \rangle$,
it is drastically altered by a measurement of only a 
tiny part of the system.
For example, 
by measurement of $\hat \sigma_z$ of the first spin, 
$| \Phi \rangle$ reduces to either
\[
| \Psi_+ \rangle \quad \mbox{when $\sigma_z = +1$},
\]
or
\[
| \Psi_- \rangle \quad \mbox{when $\sigma_z = -1$}.
\]
Hence, the results of subsequent measurements
at distant points are drastically altered 
depending on the result of the first measurement.
Note that the symmetric ground state turns into a pure-phase vacuum after the LM, 
and the symmetry is then broken. 
After that, 
the state alters only slightly by subsequent LMs, 
and the symmetry remains broken, 
because $| \Psi_\pm \rangle$ are stable against LMs.
This may be the most general mechanism of symmetry breaking in finite systems.

In summary, 
we study the stabilities of quantum states of finite macroscopic systems, 
against weak classical 
noises, weak perturbations from environments, 
and local measurements.
It is found that these stabilities are 
closely related to 
the cluster property (which describes the strength of 
spatial correlations of fluctuations of local observables)
and fluctuations of additive operators
(which are given by the sum of local operators over 
a macroscopic region).
Note that the stabilities are defined as {\em dynamical} 
properties of an {\em open}
system, whereas the cluster property and fluctuations of additive operators
are defined as {\em static} properties of a {\em closed} system.
Hence, it is non-trivial --- may be 
surprising --- that they are closely related to each other.

We thank I.\ Ojima, M.\ Ueda, H.\ Tasaki, and A.\ Ukena for discussions.
This work is partially supported by Grant-in-Aid for Scientific 
Research.


\begin{references}
\bibitem{ft}
To express and utilize the locality of the theory manifestly, 
we use a field theory throughout this paper.
\bibitem{homogeneous}
The homogeneity can be a generalized one:
Our theory is applicable to, e.g., quantum computers \cite{us}, 
in which states for some $N$ are similar to states for another $N$
because they are generated by the same algorithm.
\bibitem{ex}
In the Ising model discussed below, 
$| \Psi_+ \rangle$ and
$| \Psi_- \rangle$ become in the limit of $V \to \infty$ vectors of 
inequivalent Hilbert spaces ${\cal H}_+$ and ${\cal H}_-$, 
respectively.
Since there is no interference between them, 
$| \Phi \rangle \in {\cal H}_+ \oplus {\cal H}_-$ in this limit 
is equivalent to a classical mixture of 
$| \Psi_+ \rangle$ and
$| \Psi_- \rangle$.
\bibitem{general}
Although we here describe the case where
$f$ or $\hat f$ and $\hat a$ are real or hermitian, 
the generalization is straightforward.
\bibitem{time}
For a given $V$, this is satisfied for sufficiently small $\lambda$,
e.g., 
$1/\Gamma \sim$ 1 ms and 1 ns, respectively, for a 
NFS and an AFS with $N=10^6$, whereas $\tau_{\rm c} \sim 1$ ps.
Although $\Gamma$ varies with $\lambda$, 
we stress that 
$\Gamma$ for AFSs can be much larger than 
$\Gamma$ for NFSs
by many orders of magnitude
for any value of $\lambda$, as shown below.
\bibitem{k}
Since $f(x,t)$ for $x \not\in V$ is irrelevant, 
the spatial Fourier transform is taken over $V$ to define $g(k,\omega)$, 
hence $k$ takes discrete values with separation $\sim V^{-1/d}$.
\bibitem{relax}
When this condition is violated around $t \sim t_1$
such that $\tau_{\rm c} \ll t_1 \ll 1/\Gamma$, 
we can apply formula (\ref{r-noise-2}) again now to the evolved state 
$e^{-i \hat H t_1 }| \Psi \rangle$, 
because 
the decoherence until $t_1$ does not affects
$\Gamma$ after $t_1$ to $O(\lambda^2)$.
We may repeat this for intervals 
$[t_1,t_2), [t_2,t_3), \cdots, [t_{M-1},t_M)$ 
until some $t_M \ll 1/\Gamma$.
In cases of a rotating ferromagnet and 
a condensed state of bosons \cite{pre01}, the only quick evolution 
induced by $\hat H$ is the rotation of the direction and 
phase, respectively, of the order parameter.
In such cases, $\Gamma$ in every interval is of the same order, 
and formula (\ref{r-noise-2}) gives the correct order of magnitude
in the entire region of $\tau_{\rm c} \ll t \leq t_M$, although
$e^{-i \hat H t}| \Psi \rangle \simeq 
e^{-i \langle \hat H \rangle t}| \Psi \rangle$ 
is not satisfied for $t>t_1$.
\bibitem{like}
$(x,t_a)$ and $(y,t_b)$ are separated either time- or
space-like.
\bibitem{continuous}
The case of continuous eigenvalues
can be discussed in a similar manner.
\bibitem{intensity}
For example, electromagnetic noises at 4K
contribute only to $k=0$ component of $g(k)$ 
if the diameter of the system is less than 1cm \cite{k}.
Since $\sum_k g(k) = O(V^0)$, 
$g(0) = O(V^0)$ whereas $g(k)=O(1/V)$ 
for $k \gg 1{\rm cm}^{-1}$.
\end{references}
\end{document}